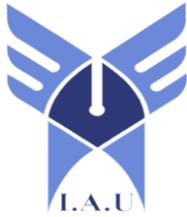

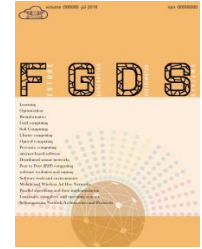

# Energy Efficient Computation Offloading and Virtual Connection Control in Uplink Small Cell Networks

Davoud Yousefi[1], Hassan Yari[1], Farzad Osouli[1], Mohammad Ebrahimi[1], Somayeh Esmalifalak[1], Morteza Johari[1], Abbas Azarnezhad[1], Fatemeh Sadeghi[1], Rogayeh Mirzapour[1]

[1] Department of Computer Engineering, Moghadas Ardabili Institute of Higher Education, Ardabil, Iran

d.yousefi.sh@gmail.com(Corresponding author)
yariarallou@gmail.com
farzadosoili@gmail.com
M.ebrahimi8676@gmail.com
somayyehesmalifalak98@gmail.com
mortez.johari@ihemardabill.acir
Azarnejad21@gmail.com
efsadeghi18@yahoo.com
rooogimirzapoor@gmail.com

**ABSTRACT**

Nowadays, the use of soft computational techniques in power systems under the umbrella of machine learning is increasing with good reception. In this paper, we first present a deep learning approach to find the optimal configuration for HetNet systems. We used a very large number of radial configurations of a test system for training purposes. We also studied the issue of joint carrier/power allocation in multilayer hierarchical networks, in addition to ensuring the quality of experience for all subscribers, to achieve optimal power efficiency. The proposed method uses an adaptive load equilibrium model that aims to achieve "almost optimal" equity among all servers from the standpoint of the key performance indicator. Unlike current model-based energy efficiency methods, we propose a joint resource allocation, energy efficiency, and flow control algorithm to solve common nonconvex and hierarchical optimization problems. Also, by referring to the allocation of continuous resources based on SLA, we extended the proposed algorithm to common flow/power control and operational power optimization algorithm to achieve optimal energy efficiency along with ensuring user's throughput limitations. Also, simulation results show that the proposed controlled power/flow optimization approach can significantly increase energy efficiency compared to conventional designs using network topology adjustment capability.

## 1. INTRODUCTION

The increasing demand for traffic in high-speed mobile networks has led researchers to plan for the next generation of wireless networks that can make significant improvements in coverage and user experience [1,2]. Improving the current load distribution system is a major issue for growing load requirements [3,4,5]. Power dissipation and resource stability is one of the most important features of the next generation of mobile systems. These characteristics mostly depend on several factors such a as distribution system expansion, load complexity and installation of distributed power generation. Several methods have been proposed in [6,7,8] to increase energy efficiency and reduce RAN loss, for example, network backhauling reconfiguration, flow control and optimal resource management and so on. Among other technical methods, network reconfiguration is a minor and insignificant solution that is done by switching the positions of serial and cross-sectional network components [9, 10]. Distribution systems have poor hybrid networks. However, work in a radial structure is recommended for simple operation and protection of such an arrangement [11, 12, 13, 14]. A mixed integer that is constrained simultaneously with linear programming is proposed in [15] and evaluated through simulations. Many metaheuristic methods are used in reconfiguration topics with separate objective functions. Genetic algorithms are widely used in the detection of optimal switching solutions with different modified forms of GA in [16, 17, 18]. Particle swarm optimization [19], harmonic search [20], bacterial search [21], and ant colony models [22] are used to solve the reconfiguration problem. In [23], the Most Probable Scenario (MPS) method was implemented to detect the least wasted configuration that deals with the most probable occurrence. Both Continuation Power Flow theorem (CPF) and load flow analysis in [24] were used to detect the maximum load capacity point. The authors in [25] proposed a new method for a dynamically reconfigured distribution system that depends on the initial topology rather than the fixed topology. [26] created the optimal radial system configuration from a hybrid system using the Kruskal algorithm. [27] used capacitor placement and network reconfiguration to minimize power loss and increase system reliability using a fuzzy approach. This paper proposes the ML method with the help of graph theory to determine the optimal network configuration. On the other hand, flow control and power optimization play an important role in energy efficiency in mobile cellular networks. In this regard, some related concepts have emerged to study the energy optimization problem in unorthodox multiple access networks to optimize network energy efficiency and increase system performance. [28] An effective game theory model has been used to allocate resources among UE's in single-carrier non-orthogonal multiple access (SC-NOMA) network to increase the efficiency of base stations. [29] Introduces a common non-optimal resource allocation and carrier allocation scheme that uses gradient decomposition and linear programming to achieve optimal weighted data rates in a multi-carrier non-orthogonal multiple access network (MC-NOMA). In [30, 31, 32, 33, 34], for a multiple-user MIMO non-orthogonal multiple access network, the authors propose a powerful source allocation algorithm to obtain maximum power under total energy consumption limits and to meet the lower boundary of data rates for edge UEs. Recently, some model-based power optimization algorithms have been proposed to increase energy efficiency or other key performance indicators in non-orthogonal multiple access networks. Resource allocation issues were investigated in [35, 36, 37] and joint power Control and resource allocation issues were studied in [38]. However, given the dynamics undeniable in next-generation wireless networks, in practical situations, achieving the formal model required in current power optimization schemes is very difficult or even impossible [39]. Some researchers have used the reinforcement learning (RL) method as an effective model less method to reduce the time complexity of the network with accessible data as input and output [40].

One of the powerful reinforcements learning methods (RL) is known as deep learning, which has been used to find an optimal solution to power optimization problems [41], which first works on training neural networks with valid input data and then receiving outputs during a real-time scenario. However, providing a sufficiently accurate data set and the optimal algorithms used for the training phase is difficult to achieve in any situation. It should be noted that the training phase itself is always time consuming. According to the above, deep learning (DL) [42, 43, 44] can be a good solution for the online decision-making process (e.g. adaptive power optimization) because in the context of reinforcement learning, the network model before starting the decision-making process for

D.Yousefi, H.Yari, F.Osouli, M.Ebrahimi, S.Esmalifalak,
M.Johari, A.Azarnezhad, F.Sadeghi, R.Mirzapour

Finding the optimal value does not require much data. However, beyond optimizing some parameters or trying to achieve convergence to an optimal value, reinforcement learning can be used to achieve a detailed decision-making strategy taking into account long-term benefits and improving network performance. It should be noted that the disadvantage of most current reinforcement learning algorithms is time-consuming convergence and these specifications make these algorithms unsuitable for large models with multiple operating spaces. Therefore, deep reinforcement learning, as a combination of deep learning and reinforcement learning, can be introduced as a suitable idea for resolving these problems. [45] Proposed an effective deep reinforcement learning algorithm called Deep Q-Learning, it is a Q system capable of using deep neural networks as a powerful tool for making decisions with the highest level of accuracy compared to conventional reinforcement learning schemes. This feature can be used to power optimization in non-orthogonal multiple access network [46], power allocation in HetNets [47] and cloud computing [48]. According to the above explanations, the main problem of Q deep learning networks is that the output is only discrete and when applying this approach to continuous scenarios (such as continuous resource allocation) we encounter quantization error. However, the unprecedented Deep Definitive Policy Gradient (DDPG) proposal introduced by [49] is capable to overcoming these issues. An in-depth definitive policy gradient is an improved version of the Definitive Policy Gradient Approach (DPG) shown in [50] that uses a functioning system to make definitive decisions and a certified system to evaluate action. The deep definitive policy gradient also provides the benefits of regular feedback and target system strategies from the definitive policy gradient to increase learning stability, which makes the deep definitive policy gradient more effective for dynamic power allocation frameworks. The research also takes into account the criteria for power distribution in hierarchical heterogeneous networks that are crucial in next-generation wireless networks and tries to propose an outstanding power sharing approach to adapt to multilayer configuration networks. We also investigate the problem of hierarchical resource allocation in an uplink NOMA system. Motivated by the above considerations, we provide a dynamic framework for improving the long-term energy efficiency of the NOMA system while ensuring the minimum data rate of all subscribers.

The first part of the framework is based entirely on the dynamic model of the deep Q learning network, while the second part combines the advantages of the definitive policy gradient and the deep definitive policy gradient according to the limitations of the load balance. Based on the DRL methods they use, we refer to this framework as continuous resource allocation based on DRL, the Continuous DRL-Based Resource Allocation Framework (CDRA).

The main idea of this paper is based on the claim that the performance of NOMA resource allocation schemes can significantly increase the joining of dynamic power optimization and random load sharing approaches. This study mainly examines the performance of a landmark resource allocation scheme with an effective load balancing approach in NOMA heterogeneous wireless networks. In this regard, we first interpret a multi-factor resource allocation plan using deep reinforcement learning. After presenting an adaptive load balance approach, we prove that the performance of the integrated plan will significantly increase the network energy efficiency, the likelihood of user QoE satisfaction, and power perspectives. The rest of this article is organized as follows. Section 2 provides a power optimized flow control in millimeter wave backhauling, in which the system model and problem formula, flow control, backhaul.

channel model and energy consumption model are discussed. Section 3 describes the proposed energy efficient flow control approach. In this section, we introduce in detail the two proposed algorithms for shared resource allocation, energy efficiency and flow control (JRAEE) and shared flow / power control, and operating power optimization (JFPCT). We also described an evaluation model for evaluating the proposed algorithms. The load sharing approach is presented in part 4, whose performance is based on deep learning approaches.

## 2. Power Optimized flow control with power in mm Backhauling waves

### 2.1. System model and formulation of problem

The system model includes a set of n base station that is shown with $BS_n\_n$ in which $N \in \{0, 1, ..., n - 1\}$ represents small base stations (as a destination relay or middle) And 0 indicators of the Macro base station as the final destination shown in Figure 1. Receptors can be identified as d = 1, 2, ..., D where N - 1 = D. Note that the downlink channel assumes two layers of orthogonal frequency division multiple access (OFDMA), so that small base stations are placed in several predefined clusters, each of which contains one cluster head.

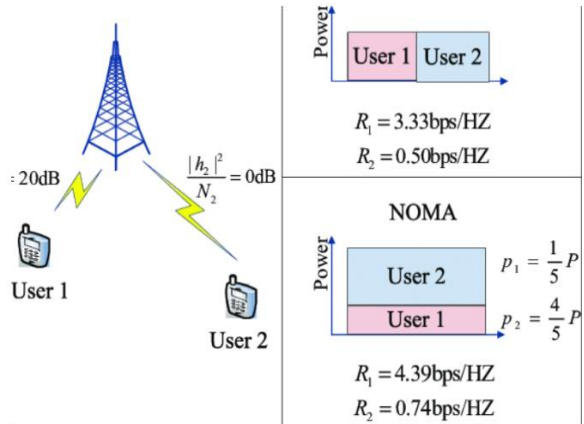

**Fig.1** The system model: user participation method and Resource allocation approach

The final destination shown in Figure 1. Receptors can be identified as d = 1, 2,...,D where N - 1 = D . Note that the downlink channel assumes two layers of orthogonal frequency division multiple access (OFDMA), so that small base stations are placed in several predefined clusters, each of which contains one cluster head. In order to limit layer interference, it is assumed that macro base stations and small base stations use different sub-channels, which are shown as $F_m$ for macro base stations and $F_s$ for small base stations. These resources are divided into sub-carriers of OFDMA. For all small base stations located in a cluster, small base stations can apply a similar sub-cell, but to remove common layer interference, there is no overlap between two small cells in this scenario. After that, we considered two adjacent macro base stations to remove intercellular interference that are far enough apart. The bandwidth carrying each macro base station is calculated as $\Delta B^m = \frac{F_m}{z_m}$ and corresponds to a small base station $\Delta B^s = \frac{F_s}{z_s}$. It is also assumed that the maximum cumulative transmission power (MTP) for a base station will be allocated equally to all of its carriers. In this scenario, the system is able to service all user J equipment specified by the $u\varepsilon \in \{1,2,...,J\}$set. Therefore, j represents the user equipment index. It is considered that each user's equipment is associated with a specific base station and adequate carriers are assigned to them. In this symbol, $u_d u_0$ show the set of user equipment associated with macro base station and small base station d, respectively. As shown in Figure 1, J is the user equipment on the network, $u_\varepsilon = \left[\bigcup_{d=1}^{D} u_d\right] \cup u_0$. Each user's equipment has a distinct data rate demand that needs to be met. The demand for user equipment $j \in u_0$ is specified as the $y_j$ and demand for user equipment $j \in u_d$ as $y_j^{(d)}$. For each small base station d, a source-destination $s^{(d)} \in \mathbb{R}^{N-1}$vector is considered where $n^{th}(n \neq d)$ with input $s_n^{(d)}$ indicates a positive flow to the network from the macro base station with the destination of the small base station d considering a single transmitter node (i.e. macro base station), $s_n^{(d)} = 0$, $\forall n \neq 0$. According to the flow control framework, the destination small cell flow can be calculated by $s_d^{(d)} \neq -s_0^{(d)}$. Based on the demand for user equipment data rate associated with the small base station d, the total demand in the small base station d is calculated as $s_d^{(d)} \equiv \Sigma_{j \in u_d} y_j^d$. The total demand for user equipment associated with the macro base station is shown as $s_0 \equiv \Sigma_{j \in u_0} y_j$

$$\mathcal{Y}_j^{(d)} = \Delta B^s \log_2\left(1 + \frac{|h_j^d|^2 P_j^{(d)}}{\sigma^2}\right), \forall_j \in u_d, \quad (1)$$

$$\mathcal{Y}_j = \Delta B^m \log_2\left(1 + \frac{|h_j|^2 P_j}{\sigma^2}\right), \forall_j \in u_0.$$

**2.2. Flow control**

In the presented framework, the one-way backhaul links are $l = 1,2,...,L$ and the occurrence matrix of the node $A \in \mathbb{R}^{N \times L}$, whose input is $A_{nl}$ assigned to the node n and the l link.

$$A_{nl} = \begin{cases} 1, & \text{n is the first node of link l} \\ -1, & \text{n is the last node of link l} \\ 0, & \text{otherwise.} \end{cases} \quad (2)$$

In equation (1), $P_j^{(d)}$ indicates the d-cell transfer power to user equipment $j \in u_d$ and $p_j$ shows the transfer macro contact power to user equipment $j \in u_0$. $|h_j^d|^2$. $|h_j^d|^2$ are channel gain indicators for $j \in u_d$ user equipment and $j \in u_0$ user equipment, respectively. $\sigma^2$ also shows the standard deviation of additive white Gaussian noise (AWGN). Note that we applied the queuing model for different QoS-Class data from the data, which is shown as picture in Figure 2.

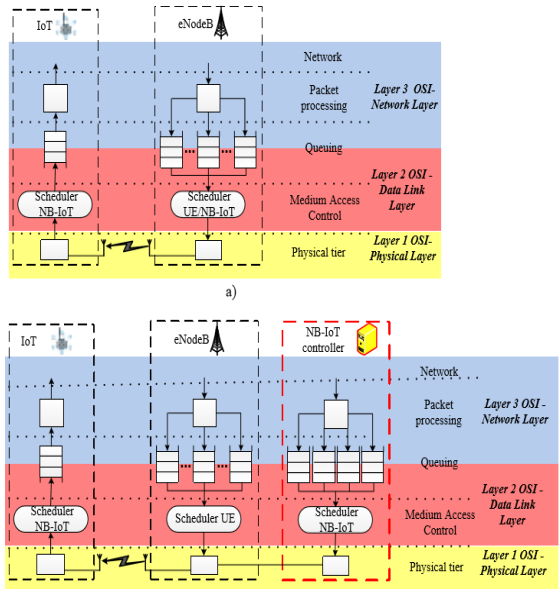

**Fig. 2** Applied SLA-based queuing model for users' throughput demands

In this scenario, O(n) shows the set of outgoing links from the n sender, and I(n) shows the set of links to the middle relay "n". We also considered $\chi_l^{(d)} \geq 0$ as flow towards the small base station d via the l link. $\chi^{(d)} \in \mathbb{R}^L$ shows the small cell flow set d. The flow control rule required in each middle relay n can be displayed as equation (3).

$$\sum_{l \in O(n)} \chi_l^{(d)} - \sum_{l \in I(n)} \chi_l^{(d)} = \begin{cases} s_0^{(d)}, n = 0 \\ 0, \forall_n \neq 0 \\ -s_n^{(d)}, n = d \end{cases} \quad (3)$$

We can also write this limit as vector matrix (4).

$$Ax^{(d)} = s^{(d)}, d = 1, 2, \ldots, D. \quad (4)$$

Considering the capacity constraints on each backhaul link l, the overall traffic flow in the l link, expressed , should be less than the radio capacity of the link $c_l$ $t_l$ the link as shown below.2.3- Backhaul channel model with millimeter wave.

$$t_l = \sum_d \chi_l^{(d)} \leq c_l. \quad (5)$$

As shown in Figure 3, mapping QoS levels to users' demands is considered by taking service classes in the proposed flow control power optimization method.

## 3. BACKHAUL CHANNEL MODEL WITH MILLIMETER WAVE

Within the millimeter wave backhaul framework, there is a unique radio access link between the cluster head and the macro base station, and several backhaul links among small base stations. For the simplicity of the issue, it is assumed that the macro base station is connected to the main network via a direct fiber. Any small contact can be connected to the macro base station via cluster head. We considered two millimeters wave sub-carriers for our backhauling network. The framework also uses 73 GHz (E-band) to backhaul the unit between macro and small contacts and 60 GHz (V-band) for several backhaul connections designed among small cells. In this scenario, the fading of transmission signals can be considered as two distinct categories: loss of sight line

$(\text{FSPL}_{(\text{dB})})$ and fading of beams related to millimeter wave loss factors $(\text{PL}_{(\text{dB})})$. These two types of emission models have already been introduced by [51].

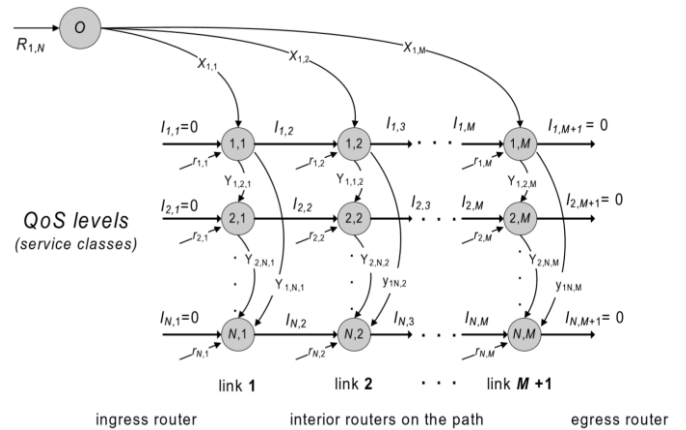

**Fig.3** Mapping QoS levels to users' demands by taking service classes.

$$\text{FSPL}_{(\text{dB})} = 92.4 + 20\log_{10}(f_{(\text{GHz})}) + 20\log_{10}(d_{(\text{km})}) \; \text{PL}_{d(\text{dB})}$$

$$d_{(\text{km})} \left( \underbrace{L_{\text{vap}} + L_{O_2}}_{\text{atmospheric gas}} + L_R \right)_{\left(\frac{\text{dB}}{\text{km}}\right)} \quad (6)$$

In this formula, d is the line of sight between source and destination and f shows frequency in GHz. $L_{o2}$, $L_{vap}$ and $L_R$ due to environmental conditions, such as oxygen, steam and rain, show a drop in the path in dB/km, respectively. Considering all the fading factors mentioned above, the total loss of path will be calculated as equation (7).

$$\text{TPL}_{(dB)} = \text{FSPL}_{(dB)} + \text{PL}_{d(dB)}. \quad (7)$$

## 4. BACKHAUL ENERGY CONSUMPTION MODEL

Based on advanced designs, there is no energy efficiency model for millimeter wave hierarchical communication. However, the use of nonlinear stochastic model to minimize total energy consumption in millimeter wave systems has already been studied in [52, 53]. This paper focuses solely on multilayer adaptive power optimization $P_l$, which is modeled on the backhauling of the l connection

$$P_l = P_{max_l}^{BH}, \quad 0 \leq P_l \leq P_{max_l}^{BH} \quad (8)$$

In which $P_{max}^{BH}$ shows the maximum transfer power in the backhaul link l, which is calculated as follows:
$$P_{max_l(dBm)}^{BH} = \text{EIRP}_{max(dBm)} + T_{xloss(dB)} - G_{Tx(dBi)} \quad 9$$

In this formula, $\text{EIRP}_{max\,(dBm)}$ represents the maximum EIRP, $T_{xloss(dB)}$ is weakening of transmission and $G_{Tx\,(dBi)}$ represents the transmitter gain. For the selected millimeter wave carriers, the radiant power equivalent to isotropicity can be achieved as [54].

V band: $\text{EIRP}_{max(dBM)} = 85_{(dBm)} - 2.x_{(dB)}$, (10)

E band: $\text{EIRP}_{max(dB\,M)} = 85_{(dBm)} - 2.y_{(dB)}$,

In this framework, we determine the difference between the transmitter's gain $G_{Tx}$ and 51dBi. y also shows the difference between the $G_{Tx}$ transmitter gain and the 50dBi. SINR at the backhaul link receiver point $l\left(\text{SINR}_{l(dB)}^{BH}\right)$ is calculated as equation (11).

$$\begin{aligned}\text{SNR}_{l(dB)}^{BH} &= P_{l(dBm)} - N_{th(dBm)} - NF_{(dB)} \\ &\quad - T_{xloss(dB)} - R_{xloss(dB)} + G_{Rx(dBi)} \\ &\quad - L_{margin} \\ &\quad - \text{TPL}_{(dB)}. \end{aligned} \quad (11)$$

In this equation, $N_{th}$ represents thermal noise and NF represents the shape of noise. After that, $G_{Rx}$ indicators, $L_{margin}$ and $R_{xloss(dB)}$ show the destination gain, connection link margin and loss of destination hardware, respectively.

In the presented framework, the one-way backhaul links are $l = 1, 2, \ldots, L$ and the occurrence matrix of the node is $A \in \mathbb{R}^{N \times L}$, whose input is $A_{nl}$ assigned to the node n and the l link.

$$EE = \frac{f_1(y,t,p)}{f_2(y,t,p)} = \frac{\sum_{j \in u_d} y_j^{(d)} + \sum_{j \in u_0} y_j}{\sum_{\forall d} \sum_{j \in u_d} \underbrace{\frac{\left(2^{\frac{y_j^{(d)}}{\Delta B^s}} - 1\right)\sigma^2}{\left|h_j^{(d)}\right|^2}}_{P_j^d} + \sum_{j \in u_0} \underbrace{\frac{\left(2^{\frac{y_j}{\Delta B^m}} - 1\right)\sigma^2}{\left|h_j\right|^2}}_{P_j} + \sum_{l \in O(n)} \underbrace{P_{max_l}^{BH} \frac{t_l}{c_l}}_{P_l}} \quad (12)$$

In this scenario, O(n) shows the set of outgoing links from the n sender, and I(n) shows the set of links to the middle relay "n". We also considered $\chi_l^{(d)} \geq 0$ as flowing towards the small base station d via the l link. $\chi^{(d)} \in \mathbb{R}^L$ shows the small cell flow set d. The flow control rule required in each middle relay n can be displayed as equation (3).

$$\begin{aligned}&\max_{t} \quad EE \\ &\text{s.t} \quad (13)\end{aligned}$$

C1: $Ax^{(d)} = s^{(d)}, d = 1,2, \ldots, D$

C2: $t_l = \sum_d x_l^{(d)}, l = 1,2, \ldots, L$

C3: $t_l \leq c_l, l = 1,2, \ldots, L$

C4: $P_l \leq P_{max_l}^{BH}, l = 1,2, \ldots, L$

C5: $x^{(d)} \geq 0, d = 1,2, \ldots, D.$

We can also write this limit as vector matrix (4).

$$\min_{t} f_2(t) \quad (14)$$

s.t  C1 ~ C5.

Taking into account the capacity constraints per backhaul link l, the overall flow of traffic in the l link expressed as $t_l$ should be less than the radio capacity of the link $c_l$ the link that is shown as below.

$$\begin{aligned}&\max_{t, y, p} \quad EE \quad (15) \\ &\text{s.t}\end{aligned}$$

C1: $s_d^{(d)} = \sum_{j \in u_d} y_j^{(d)}, d = 1,2, \ldots, D$

C2: $Ax^{(d)} = s^{(d)}, d = 1,2, \ldots, D$

C3: $t_l = \sum_d x_l^{(d)}, l = 1,2, \ldots, L$

C4: $t_l \leq c_l, l = 1,2, \ldots, L$

C5: $\sum_{\forall d}\sum_{j\in u_d}\frac{\left(2^{\frac{y_j^{(d)}}{\Delta B^S}}-1\right)\sigma^2}{\left|h_j^{(d)}\right|^2}+\sum_{l\in O(d)}P_{max_l}^{BH}\frac{t_l}{c_l}\leq P_{max}^{(d)},\forall d$

C6: $\sum_{j\in u_0}\frac{\left(2^{\frac{y_j}{\Delta B^m}}-1\right)\sigma^2}{|h_j|^2}+\sum_{l\in O(0)}P_{max_l}^{BH}\frac{t_l}{c_l}\leq P_{max}^0,$

C7: $x^{(d)}\geq 0, s^{(d)}\geq 0, \quad d=1,2,\dots,D$

C8: $\mathcal{Y}_j^{(d)}\leq \mathcal{Y}_{max}, \mathcal{Y}_j\leq \mathcal{Y}_{max},$

C9: $\mathcal{Y}_j^{(d)}\geq \mathcal{Y}_{min}, \mathcal{Y}_j\geq \mathcal{Y}_{min},$

Within the millimeter wave backhaul framework, there is a unique radio access link between the cluster head and the macro base station, and several backhauling links among small base stations. For the simplicity of the issue, it is assumed that the macro base station is connected to the main network via a direct fiber. Any small contact can be connected to the macro base station via cluster head. We considered two millimeters wave sub-carriers for our backhauling network. The framework also uses 73 GHz (E-band) to backhaul the unit between macro and small contacts and 60 GHz (V-band) for several back-hauling connections designed among small cells. In this scenario, the fading of transmission signals can be considered as two distinct categories: loss of line of sight $(FSPL_{(dB)})$ and fading of beams related to millimeter wave loss factors $(PL_{(dB)})$. These two types of emission models have already been introduced by [55].

$$\max_{t,y,p} EE = \max_{t,y,p}\frac{f_1(y,t,p)}{f_2(y,t,p)}=\eta \qquad (16)$$

s.t. C1 ~ C9.

In this formula, $d$ is the line of sight between source and destination and shows $f$ frequency in GHz. $L_{o2}$, $L_{vap}$ and $L_R$ due to environmental conditions, such as oxygen, steam and rain, show a drop in the path in dB/km, respectively. Considering all the fading factors mentioned above, the total loss of path will be calculated as equation (7).

$$\begin{cases}J_{\mathcal{Y}_{min}}\leq f_1(\mathcal{Y},t,p)\leq J_{\mathcal{Y}_{max}} \\ \left(0+\sum_{\forall j}P_j^{min}\right)\leq f_2(\mathcal{Y},t,p)\leq \left(\sum_{\forall l}P_{max_l}^{BH}+\sum_{\forall j}P_j^{max}\right)\end{cases} \qquad (17)$$

Based on advanced designs, there is no energy efficiency model for millimeter wave hierarchical communication. However, the use of nonlinear stochastic model to minimize total energy consumption in millimeter wave systems has already been studied in [56, 57]. This paper focuses solely on multilayer adaptive power optimization $P_l$ modeled on the backhauling of the $l$ connection.

$$\frac{J_{\mathcal{Y}_{min}}}{\left(\sum_{\forall l}P_{max_l}^{BH}+\sum_{\forall j}P_j^{max}\right)}\leq \eta$$
$$\leq \frac{J_{\mathcal{Y}_{max}}}{\left(0+\sum_{\forall j}P_j^{min}\right)}. \qquad (18)$$

In which $P_{max}^{BH}$ shows the maximum transfer power in the backhaul link l, which is calculated as follows:

$$\min_{t,y,p}\overline{EE} \qquad (19)$$

s.t C1 ~ C9.

In this formula, $EIRP_{max\,(dBm)}$ represents the maximum EIRP, $T_{xloss(dB)}$ is weakening of transmission and $G_{Tx\,(dBi)}$ is the transmitter gain. For the selected millimeter wave carriers, the radiant power equivalent to isotropicity can be achieved as [58].

In the presented framework, the one-way backhaul links are $l=1,2,\dots,L$ and the occurrence matrix of the node $A\in\mathbb{R}^{N\times L}$, whose input is $A_{nl}$ assigned to the node n and the $l$ link.

$$\min_{t,y,p}\alpha \qquad (20)$$

s.t C1 ~ C9.

In this scenario, O(n) shows the set of outgoing links from the n sender, and I(n) shows the set of links to the middle relay "n". We also considered $\chi_l^{(d)}\geq 0$ as flowing towards the small base station d via the l link. $\chi^{(d)}\in\mathbb{R}^L$ shows the small cell flow set d. The flow control rule required in each middle relay n can be displayed as equation (21).

find y, t, p s.t $\qquad (21)$

$f_2(y,t,p)-\hat{\alpha}f_1(y,t,p)\leq 0$

C1 ~ C9.

We can also write this limit as vector matrix (4).

$S=\{(i,j,k,m,n)\mid i\geq \cdot, i\geq \cdot, i+j+k\leq C,$
$\cdot\leq m\leq M, \cdot\leq n\leq N\} \qquad (22)$

Taking into account the capacity constraints per backhaul link $l$, the overall flow of traffic in the $l$ link expressed as $t_l$ should be less than the radio capacity of the link $c_l$ the link that is shown as below.

$$(\lambda_1 + \lambda_2 + \lambda_3)\pi_{.,.,.,.,.} = \pi_{1,.,.,.,.} + \pi_{.,1,.,.,.} + \pi_{.,.,1,.,.}$$
$$\lambda_{S_1R_1} = \cdots = \lambda_{S_NR_N} = \lambda/2 \text{ and } \lambda_{R_1D_1} = \lambda_{R_2D_2} = \cdots = \lambda_{R_ND_N} = \lambda/2$$

$$\lambda_{R.R_1} = \lambda_{R_1R_2} = \cdots = \lambda_{R_NR_{N-1}} = \lambda, \lambda = 1, \text{ and } \gamma_{th}^j = \gamma_{th}^1, j \in [1, N]. \quad (23)$$

As shown in Figure 3, mapping QoS levels to users' demands is considered by taking service classes in the proposed flow control power optimization method. Within the millimeter wave backhaul framework, there is a unique radio access link between the cluster head and the macro base station, and several backhaul links among small base stations. For the simplicity of the issue, it is assumed that the macro base station is connected to the main network via a direct fiber. Any small contact can be connected to the macro base station via cluster head. We considered two millimeters wave sub-carriers for our backhauling network. The framework also uses 73 GH (E-band) to backhaul the unit between macro and small contacts and 60 GHz (V-band) for several backhaul connections designed among small cells. In this scenario, the fading of transmission signals can be considered as two distinct categories: loss of sight line $(\text{FSPL}_{(dB)})$ and fading of beams related to millimeter wave loss factors $(\text{PL}_{(dB)})$. These two types of emission models have already been introduced by [51].

$$(i\mu_1 + \lambda_1 + j\mu_2 + \lambda_2 + k\mu_3 + \lambda_3)\pi_{i,j,k,\cdot,\cdot} =$$
$$\lambda_1 \pi_{i-1,j,k,\cdot,\cdot} + (i + 1)\mu_1 \pi_{i+1,j,k,\cdot,\cdot} + \lambda_2 \pi_{i,j-1,k,\cdot,\cdot} +$$
$$(j + 1) \mu_2 \pi_{i,j+1,k,\cdot,\cdot} + \lambda_3 \pi_{i,j,k-1,\cdot,\cdot} + (k + 1) \mu_3 \pi_{i,j,k+1,\cdot,\cdot} \quad (24)$$

In this formula, $d$ is the line of sight between source and destination and shows $f$ frequency in GHz. $L_{o2}$, $L_{vap}$ and $L_R$ due to environmental conditions, such as oxygen, steam and rain, show a drop in the path in dB/km, respectively. Considering all the fading factors mentioned above, the total loss of path will be calculated as equation (25).

$$(i\mu_1 + \alpha_1\lambda_1 + j\mu_2 + \lambda_2 + k\mu_3 + \lambda_3)\pi_{i,j,k,\cdot,\cdot} =$$
$$\alpha_2\lambda_1 \pi_{i-1,j,k,\cdot,\cdot} + (j\mu_2 + k\mu_3)\pi_{i,j,k,1,\cdot} +$$
$$\lambda_2 \pi_{i,j-1,k,\cdot,\cdot} + k\mu_3 \pi_{i,j,k,\cdot,1} + \lambda_3 \pi_{i,j,k-1,\cdot,\cdot} \quad (25)$$

Based on advanced designs, there is no energy efficiency model for millimeter wave hierarchical communication. However, the use of nonlinear stochastic model to minimize total energy consumption in millimeter wave systems has already been studied in [56, 57]. This paper focuses solely on multilayer adaptive power optimization $P_l$ modeled on the backhauling of the $l$ connection.

$$\begin{pmatrix} \alpha_5 i\mu_1 + \alpha_1\lambda_1 + \alpha_v(1 - \alpha_5)i\mu_1 + \\ \alpha_5 j\mu_2 + \alpha_v(1 - \alpha_5)j\mu_2 + \alpha_8 k\lambda_1 + \\ \alpha_5 k\mu_3 + \alpha_v(1 - \alpha_5)k\mu_3 + \alpha_9\lambda_3 \end{pmatrix} \pi_{i,j,k,m,n}$$
$$= \alpha_2\lambda_1 \pi_{i-1,j+1,k,m-1,n}$$

$$+\alpha_2\lambda_1 \pi_{i-1,j+1,k,m,n-1} + \alpha_2\lambda_1 \pi_{i,j,k,m-1,n} + \alpha_2\lambda_1 \pi_{i,j,k,m,n-1} \quad (26)$$

In which $P_{max}^{BH}$ shows the maximum transfer power in the backhaul link $l$, which is calculated as follows:

$$\pi Q = 0; \text{ and } \sum_{\forall S} \pi_{i,j,k,m,n=1} \quad (27)$$

In this formula, $EIRP_{max\ (dBm)}$ represents the maximum EIRP, $T_{xloss(dB)}$ is weakening of transmission and $G_{Tx\ (dBi)}$ is the transmitter gain. For the selected millimeter wave carriers, the radiant power equivalent to isotropicity can be achieved as [58].

In the presented framework, the one-way backhaul links are $l = 1,2,...,L$ and the occurrence matrix of the node $A \in \mathbb{R}^{N \times L}$, whose input is $A_{nl}$ assigned to the node $n$ and the $l$ link.

$$PF_1 = \sum_{\substack{\forall S \\ i+j+k=C \text{ and } k=0}} \frac{\lambda_1 j}{(C-j)\lambda_2(1-PB_2)} \pi_{i,j,k,m,n} \quad (28)$$

$$PF_2 = \sum_{\substack{\forall S \\ i+j+k=C \text{ and } n=N}} \frac{\lambda_1 k}{(C-j)\lambda_3(1-PB_2)} \pi_{i,j,k,m,n} \quad (29)$$

In this scenario, O(n) shows the set of outgoing links from the n sender, and I(n) shows the set of links to the middle relay "n". We also considered $\chi_l^{(d)} \geq 0$ as flowing towards the small base station d via the $l$ link. $\chi^{(d)} \in \mathbb{R}^L$ shows the small cell flow set d. The flow control rule required in each middle relay n can be displayed as equation (30).

$$SCR_1 = \sum_{\forall S} j\mu_2 \pi_{i,j,k,m,n} \quad (30)$$

$$SCR_2 = \sum_{\forall S} j\mu_3 \pi_{i,j,k,m,n} \quad (31)$$

**D.Yousefi, H.Yari, F.Osouli, M.Ebrahimi, S.Esmalifalak, M.Johari, A.Azarnezhad, F.Sadeghi, R.Mirzapour**

We can also write this limit as vector matrix (4).

$$U = \sum_{\forall S} \frac{i+j+k}{C} \pi_{i,j,k,m,n} \quad (32)$$

Taking into account the capacity constraints per backhaul link $l$, the overall flow of traffic in the $l$ link expressed as $t_l$ should be less than the radio capacity of the link $c_l$ the link that is shown as below.

$$D_1 = \frac{L_1}{\lambda_1} \quad (33)$$

Which

$$L_1 \sum_{\substack{\forall S \\ \cdot \leq m \leq M}} m\pi_{i,j,k,m,n}$$

As shown in Figure 3, mapping QoS levels to users' demands is considered by taking service classes in the proposed flow control power optimization method. Within the millimeter wave backhaul framework, there is a unique radio access link between the cluster head and the macro base station, and several backhaul links among small base stations. For the simplicity of the issue, it is assumed that the macro base station is connected to the main network via a direct fiber. Any small contact can be connected to the macro base station via cluster head. We considered two millimeters wave sub-carriers for our backhauling network. The framework also uses 73 GHz (E-band) to backhaul the unit between macro and small contacts and 60 GHz (V-band) for several backhaul connections designed among small cells. In this scenario, the fading of transmission signals can be considered as two distinct categories: loss of sight line $(FSPL_{(dB)})$ and fading of beams related to millimeter wave loss factors $(PL_{(dB)})$. These two types of emission models have already been introduced by [51].

$$D_2 = \frac{L_2}{\lambda_3 + R_{Interrupt}} \quad (34)$$

In which,

$$L_1 = \sum_{\substack{\forall S \\ 0 \leq n \leq N}} n\pi_{i,j,k,m,n} \quad \text{and} \quad R_{Interrup}$$

$$= \sum_{\substack{\forall S \\ i+j+k=C \\ v<C}} \frac{\lambda_1 k}{C-v} \pi_{i,j,k,m,n}$$

In this formula, $d$ is the line of sight between source and destination and shows $f$ frequency in GHz. $L_{o2}$, $L_{vap}$ and $L_R$ due to environmental conditions, such as oxygen, steam and rain, show a drop in the path in dB/km, respectively. Considering all the fading factors mentioned above, the total loss of path will be calculated as equation (7).

$$MRT_i(t) = \frac{1}{\beta_{k,m}(t)\mu_i - \lambda_i(t)}, \quad (35)$$

Based on advanced designs, there is no energy efficiency model for millimeter wave hierarchical communication. However, the use of nonlinear stochastic model to minimize total energy consumption in millimeter wave systems has already been studied in [56, 57]. This paper focuses solely on multilayer adaptive power optimization $P_l$ modeled on the backhauling of the $l$ connection.

$$\hat{s}_i(t+1) = \hat{s}_i(t) + \alpha(s_i(t) - \hat{s}_i(t)). \quad (36)$$

In which $P_{max}^{BH}$ shows the maximum transfer power in the backhaul link l, which is calculated as follows:

$$\hat{s}_i(t+1) = \hat{s}_i(t) \text{ for } j \neq i, j \in [1,n] \quad (37)$$

In this formula, $EIRP_{max\,(dBm)}$ represents the maximum EIRP, $T_{xloss(dB)}$ is weakening of transmission and $G_{Tx\,(dBi)}$ is the transmitter gain For the selected millimeter wave carriers, the radiant power equivalent to isotropicity can be achieved as [58].

## 5. Simulation Result

Figure 4 compares the average throughput of the proposed approach, Energy Efficient Computation Offloading and Virtual Connection Control in Uplink Small Cell Networks (EECO-VC) and CN-SWIPT scheme with equal maximum transmission power. Based on this figure, it is obvious that the average sum data rate increases with increasing the signal to noise ratio. It can also be seen that the EECO-VC algorithm performs much better than other algorithms in terms of higher data rate. Because the EECO-VC algorithm has the required flexibility to dedicate resources to the network entities.

| Parameter | Value |
|---|---|
| Configuration of the Network | Mobile Network, X-sectored BSs |
| sensor distribution model | uniform (U) and hotspot (Hs), |
| transmit backoff | 1.5 dB |
| Base Station MTP | 43 dBm |
| Codec strategy | Adaptive multi-rate |
| $Rx\ loss$ & $Tx\ loss$ | 3 dB |
| Propagation model | Okumura-Hata |
| Fairness Index | Security/ Throughput |
| Upper bound of iteration | 2000 |
| $L_{margin}$ | 5 dBm |
| Learning factor $c_1 = c_2$ | 1.1 |
| Weighting factor $\omega_{max}$ MAX | 0.77 |
| Weighting factor $\omega_{min}$ MIN | 0.28 |
| MAX Sensor Power $P_m^o$ | 90 w |
| MIN Sensor Power $P_s^o$ | 5 w |
| Sensor transmit range | 30 m |

**Table 1**. Main implementation factors

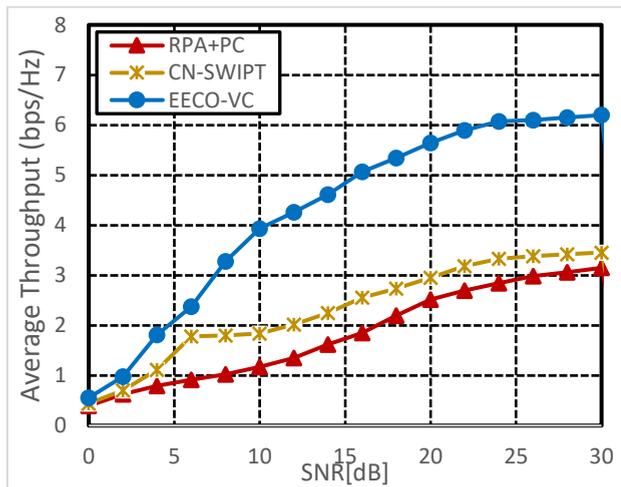

**Fig. 4** Average throughput vs. signal to noise ratio

This trend decreases slightly with an increase in N, because the algorithm reduces the throughput available to each of the nodes. In contrast, the demand for the throughput of each node is the same in all random power allocation, equal power allocation, CN-SWIPT algorithms, because they all provide the minimum throughput for each N. The throughput decreases exponentially with increasing number of N. Because with increasing N, the demand for data rate decreases. Because, the same MTP is shared equally between the nodes.

Based on the achieved results in Figure 5, the average sum-rate increases almost linearly with increasing N in the EECO-VC algorithm. While all three other two algorithms, CN-SWIPT and RPA/PC show slight improvement.

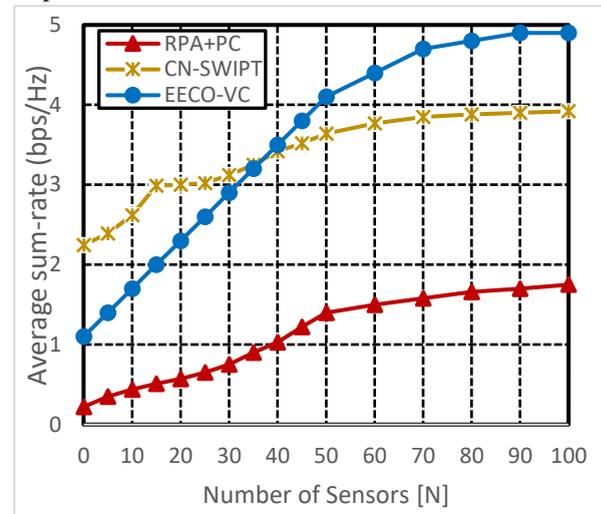

**Fig. 5** Average sum rate vs. number of sensors

Figure 6 shows the capacity of backhaul links and their average traffic (link usage in percentage) for random power allocation, MSB-MSN and CN-SWIPT. Based on this figure, it can be seen that the EECO-VC algorithm has the best performance in terms of load balancing and link usage and capacity. So, it has the highest possible efficiency in using backhaul links. Also, the high capacity of backhaul links reduces the potential for the backhaul link to be trapped in the bottleneck while sending the traffic flow to the central network.

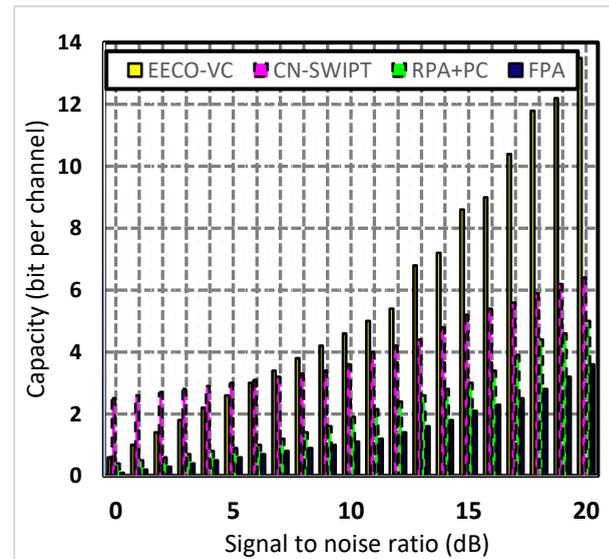

**Fig. 6** Backhaul links capacity vs. signal to noise ratio

## 6. Conclusion

This paper presents a novel approach to Energy Efficient Computation Offloading and Virtual Connection Control in Uplink Small Cell Networks (EECO-VC), which not only increases the robustness of the base stations' functionality but also maximizes the network energy efficiency in the distributed beamforming by efficient user association and resource control. In this paper, when the $\mathcal{L}_i$ link cost of the EECO-VC algorithm is used, the mobile sensor network maintains its energy efficiency level significantly more than in a state where power control techniques are not used. In future studies, more EECO-VC energy consumption should be evaluated in wireless mobile networks considering non-ideal cooperative beamforming conditions so that information needs to be transmitted frequently. In future, we plan to explore the potential of multi-agent smart queuing in various HetNet scenarios such as privacy-aware recommendation and store cell recommendation. We will also plan to examine how to exploit multi-modal data in the mobile sensor networks to further improve the proposed model.